\documentclass[preprint, 12pt, final]{elsarticle}

\usepackage{epsfig}
\usepackage{graphicx}
\usepackage{subcaption}

\usepackage{hyperref}
\hypersetup{
     colorlinks=true,
     linkcolor=blue,
     filecolor=blue,
     citecolor = black,      
     urlcolor=cyan,
     }
\usepackage{url}

\usepackage{amssymb}
\usepackage{amsthm}
\usepackage{amsmath}

\theoremstyle{plain}
\newtheorem{theorem}{Theorem}[section]

\theoremstyle{definition}

\usepackage{booktabs} 
\usepackage{colortbl}
\usepackage[table]{xcolor}

\usepackage{tcolorbox}

\usepackage{lineno,hyperref}
\journal{Advances in applied mathematics}

\begin{document}

\begin{frontmatter}

\title{Estimating the impact of non-pharmaceutical interventions and vaccination on the progress of the COVID-19 epidemic in Mexico: a mathematical approach }


\author[address2]{{Hugo Flores-Arguedas}}
\ead{hugo.flores@cimat.mx}

\author[address3]{Jos\'e Ariel Camacho-Guti\'errez}

\author[address1]{Fernando Salda\~na}

\address[address2]{Centro de Investigaci\'{o}n en Matem\'{a}ticas, 36023 Guanajuato, Guanajuato, Mexico}

\address[address3]{Facultad de Ciencias, Universidad Aut\'onoma de Baja California, 22860 Baja California, Mexico}

\address[address1]{Instituto de Matem\'aticas, Campus Juriquilla, 76230, Universidad Nacional Aut\'onoma de M\'exico, Qu\'eretaro, Mexico}

\begin{abstract}
Non-pharmaceutical interventions have been critical in the fight against the COVID-19 pandemic. However, these sanitary measures have been partially lifted due to socioeconomic factors causing a worrisome rebound of the epidemic in several countries. In this work, we assess the effectiveness of the mitigation implemented to constrain the spread of SARS-CoV-2 in the Mexican territory during 2020. We also investigate to what extent the initial deployment of the vaccine will help to mitigate the pandemic and reduce the need for social distancing and other mobility restrictions. Our modeling approach is based on a simple mechanistic Kermack-McKendrick-type model. To quantify the effect of NPIs, we perform a monthly Bayesian inference using officially published data. The results suggest that in the absence of the sanitary measures, the cumulative number of infections, hospitalizations, and deaths would have been at least twice the official number. Moreover, for low vaccine coverage levels, relaxing NPIs may dramatically increase the disease burden; therefore, safety measures are of critical importance at the early stages of vaccination. The simulations also suggest that it may be more desirable to employ a vaccine with low efficacy but reach a high coverage than a vaccine with high effectiveness but low coverage levels. This supports the hypothesis that single doses to more individuals will be more effective than two doses for every person.

\end{abstract}


\end{frontmatter}

\section{Introduction}

Since the beginning of the pandemic, the scientific community has acted fast to better understand several aspects of COVID-19, including epidemiological, biological, immunological, and virological features. Mathematical modeling has been crucial in helping public health officers make informed decisions \citep{roda2020difficult}. In particular, there is a growing literature on epidemiological modeling papers that have been mainly used to forecast the epidemic dynamics in specific countries or cities, see, for example, \cite{aguiar2020modelling, bertozzi2020challenges, ku2020epidemiological, maier2020effective, petropoulos2020forecasting, ribeiro2020short, sarkar2020modeling, zhao2020staggered}. Mathematical models have also been central to evaluate the procedures involved in the containment of the pandemic \cite{anirudh2020mathematical}. Many governments worldwide have implemented national lockdowns as extreme measures to stop disease spread. Lockdowns in addition to other non-pharmaceutical interventions (NPIs) such as mask-wearing, social distancing, temperature screening, closure of schools, restaurants, bars, and other places for social gathering have been of paramount importance in the fight against the pandemic. However, although such measures have had a significant impact in reducing the number of deaths and infections, helping to decrease the risk of health services being overwhelmed, their cost to society and economic life have been huge  \cite{mandel2020economic}. Hence, public health authorities are constantly monitoring the current state of the epidemic to evaluate lockdown exit strategies, and once again the use of mathematical models becomes a valuable tool to study the impact of partial mobility restrictions and the optimal time to relax the imposed restrictions \cite{aguiar2020modelling}.

There have been many modeling efforts presented to analyze and understand the COVID-19 pandemic in Mexico \cite{acuna2020modeling, capistran2020forecasting, hernandez2020host, mena2020using, saldana2020modeling, saldana2020trade, santamaria2020possible, santana2020lifting, torrealba2020modeling}.  The study of the early phase of the pandemic together with estimations of the basic and effective reproduction numbers is presented in \cite{acuna2020modeling, mena2020using, saldana2020modeling}. In \cite{capistran2020forecasting}, the authors present a forecasting model aiming to predict hospital occupancy. Using both hospital admittance confirmed cases and deaths, they infer the contact rate and the initial conditions of the dynamical system, considering breakpoints to model lockdown interventions and the increase in effective population size due to lockdown relaxation.  In \cite{santana2020lifting}, the authors use a mathematical model to characterize the impact of short duration transmission events. They showed that super-spreading events have been one of the main drivers of the epidemic in Mexico.

Beyond NPIs, there has been a vast-scale effort by researchers and pharmaceutical companies to develop an effective and safe vaccine to prevent infection with the SARS-CoV-2 and now several vaccines have been approved by national regulatory authorities. The availability of a vaccine represents an important positive step towards the control of the pandemic and the hope to return to normality. Nevertheless, the implementation of mass vaccination worldwide involve several financial, logistic, and social challenges \cite{su2021vaccines}. Moreover, even for a very effective vaccine, the immunization coverage needed to reach herd immunity levels and successfully control the pandemic may be very high and potentially difficult to achieve.

The first goal of this work is to evaluate the impact of the sanitary measures implemented in the mitigation of the COVID-19 pandemic in Mexico during the year 2020. We also investigate to what extent the initial deployment of the vaccine will help to mitigate the pandemic and reduce the need for social distancing and other mitigation measures. To this end, we fit a simple mechanistic Kermack-McKendrick-type model using official data of the COVID-19 epidemic in Mexico. We use Bayesian inference to calibrate the state variables and estimate how key parameters have been changing alongside the epidemic. As in \cite{capistran2020forecasting}, our modeling approach assumes that as lockdown measures as relaxed, more individuals become in contact with the outbreak. In other words, lockdown-relaxations not only cause a change in the transmission rates but also causes changes in the effective size of the population at risk.

The rest of the paper is structured as follows. In the next section, we present the mathematical model and use officially published data on the daily number of confirmed cases, hospitalizations, and cumulative deaths during the year 2020 to perform a monthly parameter inference. In Section \ref{sec:Retrospective}, such results are used to assess the role of NPIs in the mitigation of the COVID-19 pandemic. We also explore several vaccination scenarios depending on the immunization coverage, delivery time, and vaccine efficacy. A discussion of the results is presented in Section \ref{sec:discussion}.

\section{Methods}
\subsection{Model formulation}
The model presented here is based on the mathematical model first introduced in a previous work  \cite{saldana2020trade}. The model is an extension of the classical SEIR Kermack-McKendrick-type model tailored to incorporates the most important features of the COVID-19 disease and the population-level impact of vaccination.\par 
The disease dynamics are described by the following system of differential equations
\begin{equation}
\begin{aligned}
S' &= - \lambda S - \phi S,\\
V' &= -(1-\psi )\lambda V + \phi S,\\
E' &= \lambda S - k E,\\
\tilde{E}' &= \lambda (1-\psi )V  - k \tilde{E},\\
A' &= (1-p)k E  + (1-\tilde{p})k \tilde{E}  - \gamma_{A}A,\\
I' &= p k E +\tilde{p}k\tilde{E} - \gamma I - \eta I -\mu I,\\
H' &= \eta I - \gamma_{H}H - mH,\\
R' &= \gamma_{A}A + \gamma I + \gamma_{H} H,\\
D' &= \mu I + m H,\\
\end{aligned}
\label{model1}
\end{equation}
where $S, V, E, \tilde{E}, A, I, H$ and $R$ represent the number of susceptible, vaccinated susceptible, exposed, vaccinated exposed, asymptomatic infectious, symptomatic infectious, hospitalized, and recovered individuals, respectively. In this study, the asymptomatic class $A$ includes infected individuals with no symptoms but also considers mild symptomatic infections. The symptomatic class $I$ consists of individuals who develop severe disease and therefore are expected to have fewer contacts in comparison with individuals in the $A$ class. Considering disease-induced deaths, the total population size, denoted $N(t)$, is $N(t)=S(t)+V(t)+E(t)+\tilde{E}(t)+A(T)+I(T)+H(t)+R(t)+D(t)$. \par
The force of infection (FOI) $\lambda=(\beta_{A}A + \beta_{I}I)/N$ in model \eqref{model1} represents the classical standard incidence, where $\beta_{A}$ and $\beta_{I}$ are the effective contact rates for the asymptomatic and symptomatic infectious classes, respectively. Note that we are assuming that the hospitalized class is effectively isolated and does not contribute to the FOI. After a mean latent period of $1/k$, a proportion $p$ of the exposed class $E$ transition to the symptomatic infectious class $I$, while the other proportion $1-p$ enter the asymptomatic infectious class $A$. The parameters $\gamma_{A}$, $\gamma$, and $\gamma_{H}$ are the recovery rates of the classes $A$, $I$, and $H$, respectively. The parameter $\eta$ denotes the rate of transition from the $I$ class to the hospitalized class $H$. Individuals in the symptomatic and hospitalized classes experience disease-induced death at rates $\mu$ and $m$, respectively. We incorporate vaccination in the susceptible class with a rate $\phi$. In our model, vaccination not only prevents SARS-CoV-2 infection but also prevents severe symptomatic COVID-19 disease. The parameter $\psi$ is the vaccine efficacy to prevent infection and $1-\tilde{p}$ is the fraction of vaccinated individuals who after infection do not develop severe disease. Therefore, $1-\tilde{p}$ may be interpreted as the vaccine efficacy to prevent severe disease. The basic qualitative properties of the model \eqref{model1} and the reproduction numbers are studied in \ref{sec:appendixA}. \par 

\subsection{Data, parameter estimates and Bayesian inference}
\label{sec:parameters}
In this section, we perform a parameter inference using a Bayesian approach. Our main objective is to estimate the impact of the sanitary emergency measures implemented to control the spread of the virus. However, such measures have been relaxed or only partially implemented alongside the epidemic depending on several circumstances. The initial phase of the epidemic in Mexico covers from February 17 to March 22, 2020. On March 23, 2020, phase 2 was declared, which primarily includes the suspension of certain non-essential economic activities, the restriction of massive congregations, and the recommendation of home quarantine to the general population. On March 30, 2020, a sanitary emergency was declared and the public health authorities implemented a national lockdown until May 31, 2020. After this, the lockdown was lifted and other mitigation measures were only partially implemented. These changes have had a significant impact on the value of the transmission parameters \cite{capistran2020forecasting, saldana2020modeling}. We take this into account making a parameter inference by periods.
We consider the estimation of parameters by month using three sets of data obtained from the daily report of the Mexican Federal Health Secretary  \cite{datosSS}: (i) new daily reported infections, (ii) new daily hospitalizations, and (iii) cumulative deaths in Mexico. We remark that this data corresponds to the confirmed cases on the date that the patient approached the medical center and not on the day its symptoms began. Moreover, the testing rate in Mexico is the lowest among the OECD countries \cite{mena2020using}, so the data on the confirmed cases corresponds to symptomatic infections, see \ref{sec:appendixB} for the details of the inference. 

\begin{figure}
\centering
\includegraphics[height=0.7	\textheight]{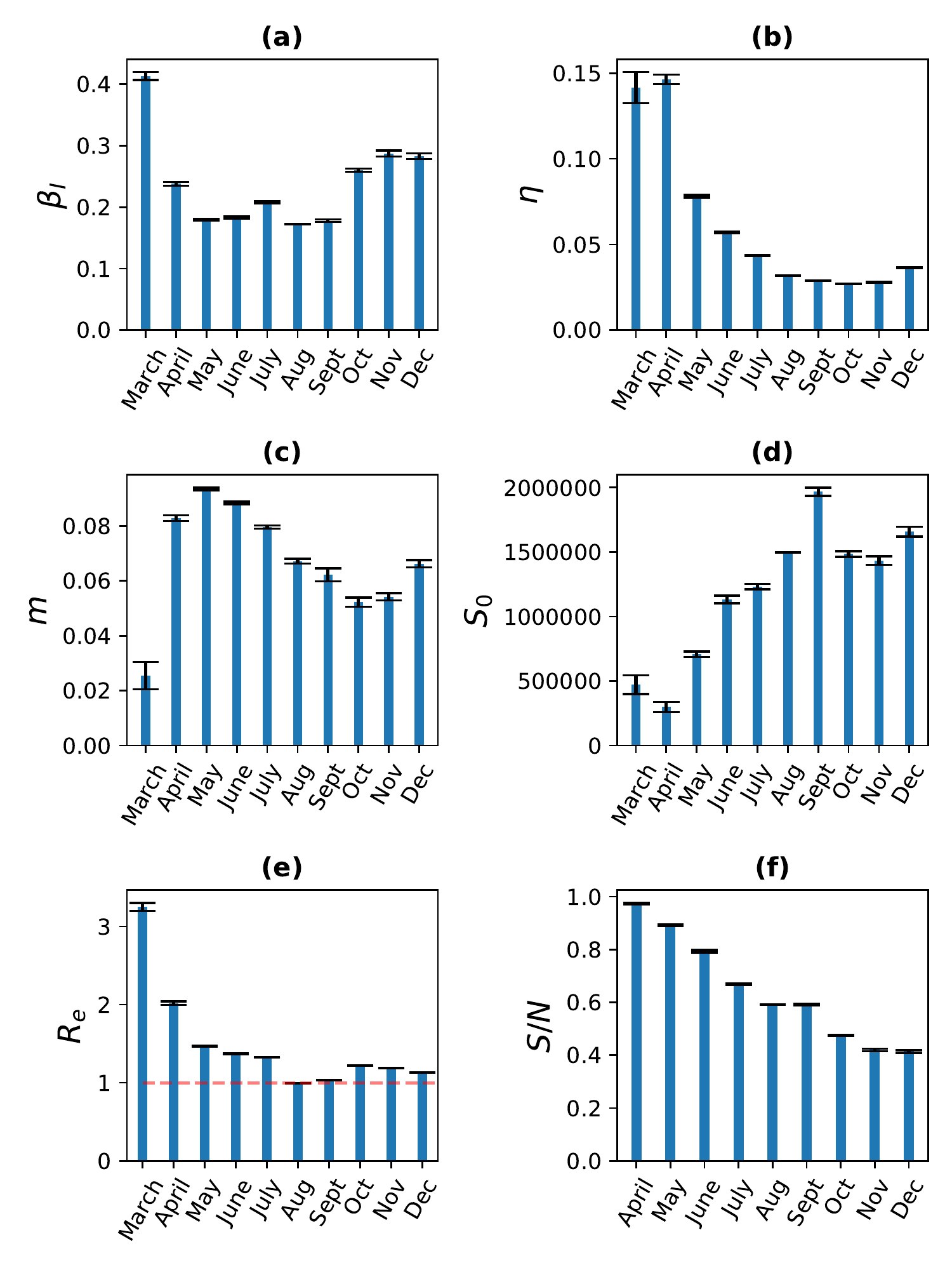}
\caption{Mean and variance (vertical black lines) of (a) the effective contact rate of the symptomatic class $\beta_I$, (b) the hospitalization's rate $\eta$, (c) the mortality rate from hospitalized patients $m$, (d) the effective susceptible population $S_0$, (e) the effective reproduction number, and (f) the susceptible fraction .}
\label{fig1}
\end{figure}

As in \cite{capistran2020forecasting, ku2020epidemiological, maier2020effective, mena2020using}, we assume that partial lifting of lockdowns and NPIs not only affects the transmission rate but also the effective population size, $N_e$, that usually satisfies $N_e \ll N$. We remark that deterministic Kermack-McKendrick-type models predict a single epidemic wave. However, our modeling approach allows us to capture the possibility of multiple waves induced by NPIs relaxation. Hence, for the inference, we consider $N_e = S_0+E_0+A(0)+I_0+H(0)+R(0)+D(0)$. The vector of parameter to estimate in the Bayesian formulation is $\mathbf{x}= (S_0, E_0, I_0,\beta_I, \eta, m)$.  The inference process is performed in a monthly way from March to December (see Figure \ref{fig1}).

\begin{figure}[h!]
\begin{subfigure}[h]{1\textwidth}
 \includegraphics[width=\textwidth]{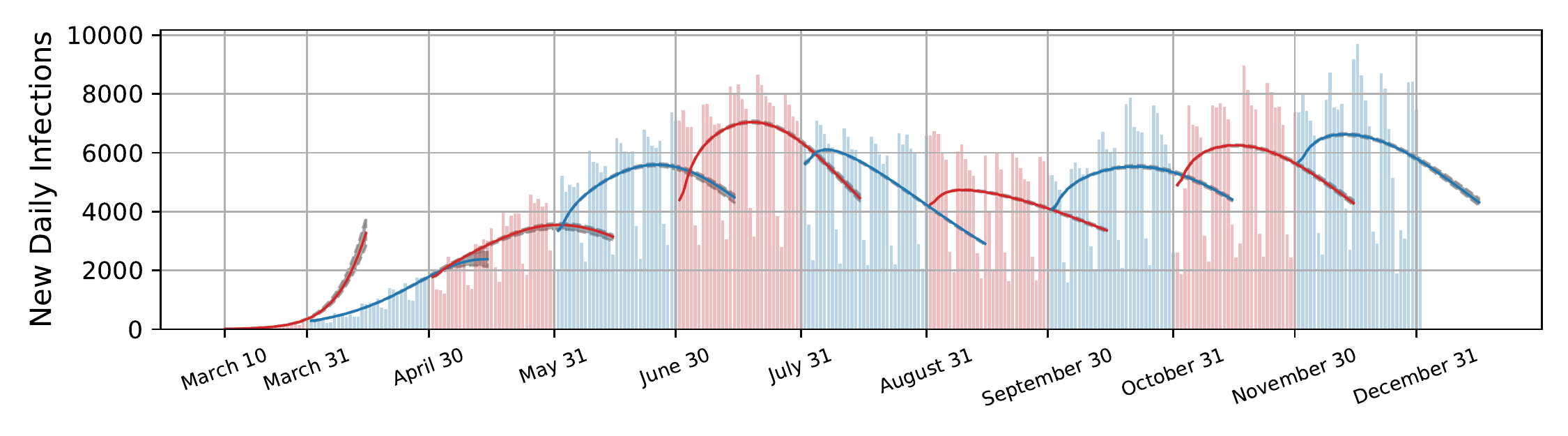}\caption{ }
\end{subfigure} 
 \begin{subfigure}[h]{1\textwidth} 
 \includegraphics[width=\textwidth]{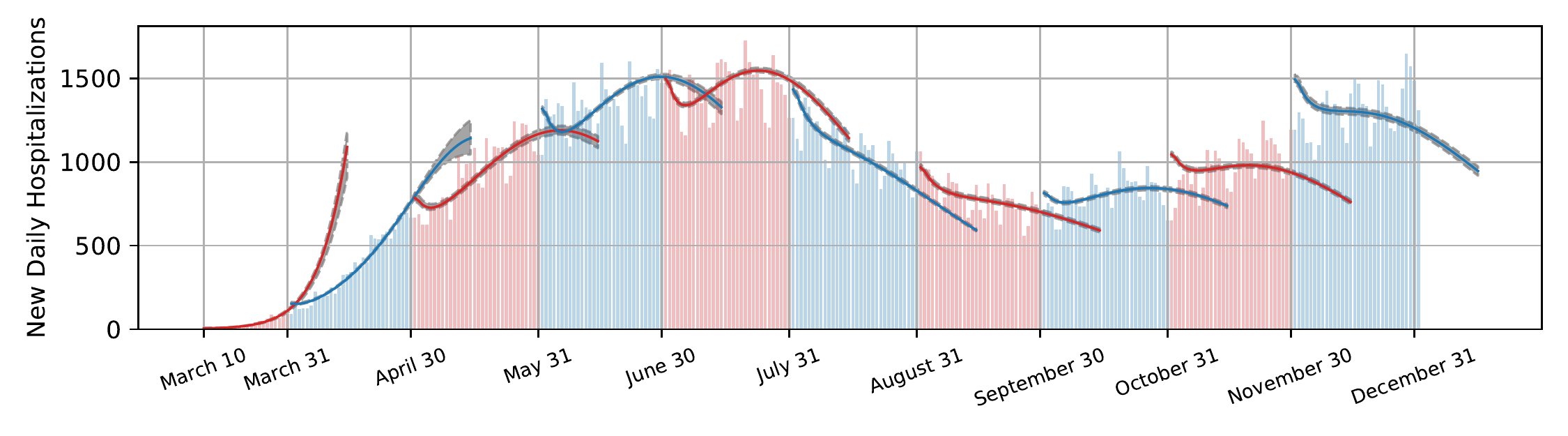}\caption{ }
 \end{subfigure}
  \begin{subfigure}[h]{1\textwidth} 
 \includegraphics[width=\textwidth]{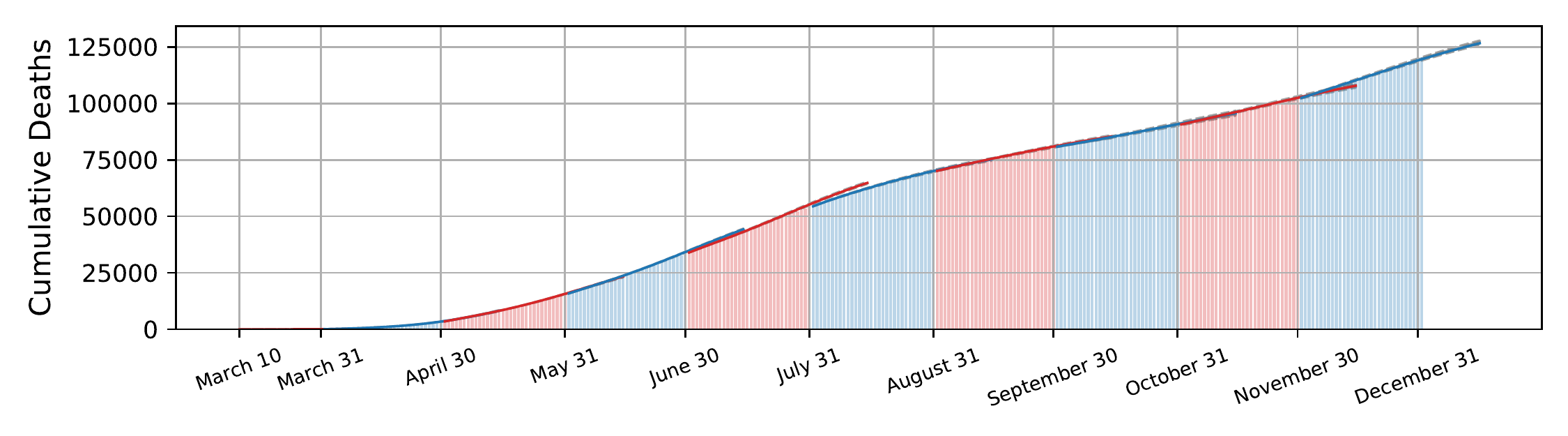}\caption{ }
 \end{subfigure}
\caption{Monthly estimation for new daily infections (a), new daily hospitalizations (b) and cumulative deaths (c) in the Mexican population. Vertical lines represent official data while solid lines represent our model predictions for the respective month. Observe that we extended our model predictions to overlap some unseen data of the next month.}
\label{fig2}
\end{figure}

As expected, the highest value of the effective contact rate in the symptomatic class is reached at the beginning of the pandemic (March), before the national lockdown and NPIs implementation. The maximum for the hospitalization's rate is reached in April, one month after the start of the pandemic in Mexico, and the maximum in the mortality rate from hospitalized patients reached in May, respectively (see Figure \ref{fig1}). At the beginning of the pandemic, the implementation of the lockdown in Mexico produced a small effective susceptible population for March and April. The progression of the pandemic and the relaxation of the measures produced a monthly increase until September. After this, an increase in the new daily infections and the contact rates in October and November produced measures that resulted in a decrease in the corresponding $S_0$. In December a new increase is noted probably due to the end of the year festivities. Moreover, observe that the effective reproduction number $\mathcal{R}_{e}$ was very high in the early phase of the epidemic. The national lockdown reduced significantly the value of $\mathcal{R}_{e}$ achieving its lowest value in August. After lockdown-relaxation, June 1, the value of $\mathcal{R}_{e}$ has been oscillating close to unity. The susceptible fraction $S/N$ is practically decreasing along the period from April to December. This behavior was broken in September, same month where the effective susceptible population reaches its maximum value.  Performing a parameter inference by periods allow us to obtain accurate estimations on the number of daily infections, hospitalizations, and cumulative deaths and make a reliable short-term prediction of such outcomes per period (see Figure \ref{fig2}).

\section{Results}
\subsection{Retrospective evaluation of non-pharmaceutical interventions and vaccination impact}
\label{sec:Retrospective}

The national lockdown in Mexico was officially lifted on May 31. However, after this date, NPIs were issued through mass media by public health authorities and are still partially implemented in the population. In this section, we investigate the impact of such NPIs in the control of the transmission dynamics of SARS-CoV-2 for the first year of the pandemic in Mexico. For this, we compare with a theoretical case in which no mitigation measures were implemented. We must remark that after the national lockdown was lifted, NPIs were state-specific in Mexico. Nevertheless, for simplicity, we are considering the total data for the whole country.

\begin{figure}[h!]
\begin{subfigure}[h]{0.99\textwidth}
 \includegraphics[width=1\textwidth]{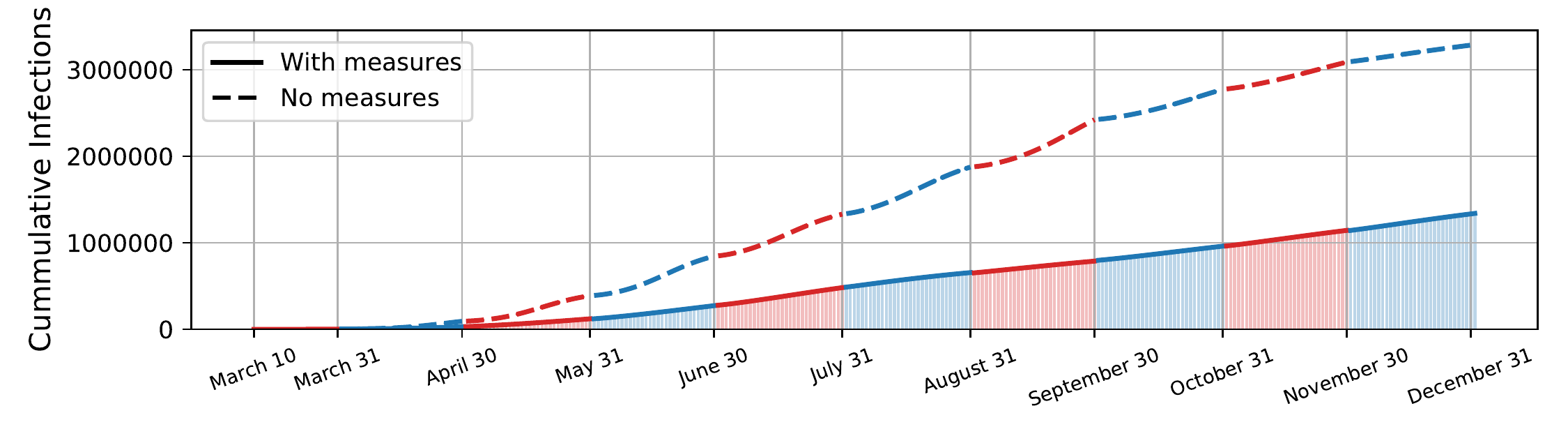}\caption{ }
\end{subfigure} 
 \begin{subfigure}[h]{0.99\textwidth} 
 \includegraphics[width=1\textwidth]{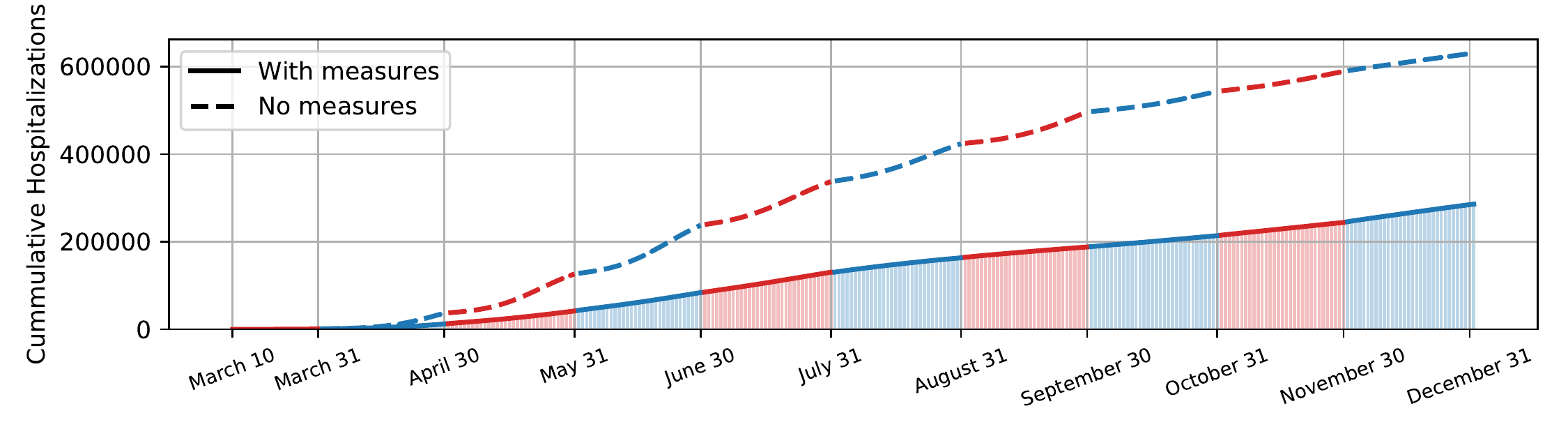}\caption{ }
 \end{subfigure}
  \begin{subfigure}[h]{0.99\textwidth} 
 \includegraphics[width=1\textwidth]{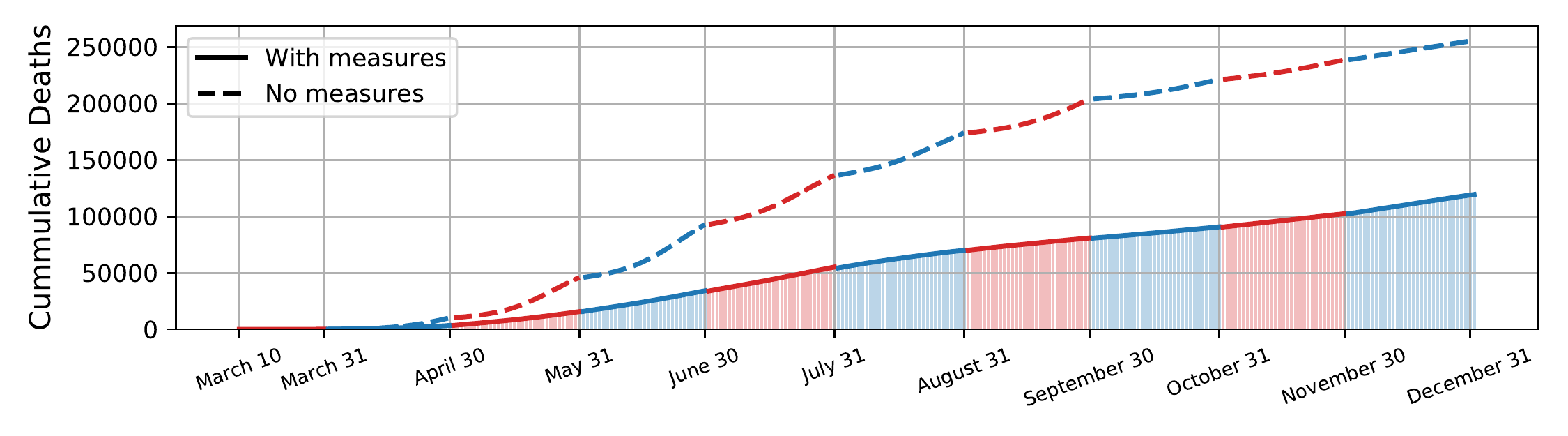}\caption{ }
 \end{subfigure}
\caption{Monthly worst-case scenario for cumulative number of infections (a), hospitalizations (b) and deaths (c) predicted in the absence of NPIs or any sanitary measures (dotted lines). Vertical lines represent official published data by the Mexican Secretary of Health since the beginning of the pandemic until December 30, 2020.}
\label{fig3}
\end{figure}

In Fig. \eqref{fig3} we evaluate a counterfactual scenario that reflects a monthly worst-case scenario, in which no sanitary measures were implemented during the whole year of 2020. For each month, the no-measures scenario was obtained by using the same initial conditions from the Bayesian inference, but changing the value of the estimated $\beta$ for that month to the value of $\beta$ at the beginning of the pandemic. Such simulations allow us to quantify the impact of NPIs in the reduction of the burden caused by COVID-19 in the Mexican population. The outcomes of interest are the cumulative number of infections, hospitalizations, and deaths predicted in the worst-case scenario in comparison with the official data. The results suggest that in the absence of the sanitary measures, the cumulative number of infections by the end of 2020 would have been above 3 million cases which is more than twice the official number. We observe a similar pattern for the cumulative number of hospitalizations and deaths, that is, they would have presented at least a two-fold increase in the absence of NPIs. From these results, it is evident that the implementation of NPIs has been of paramount importance to reduce the transmission of SARS-CoV-2 and the morbidity caused by COVID-19 disease.

\subsection{Impact of vaccine introduction}
\label{sec:Vaccination}

Since a vaccination program is starting in Mexico, we would like to know if there are vaccination scenarios that allow similar savings to those obtained by the implementation of NPIs. Results from leading vaccine developers have shown that their vaccines are more than 90\% effective to prevent infection with SARS-CoV-2 (see \cite{saldana2020trade} and the references therein). Nevertheless, the success of a vaccination program depends not only on the vaccine efficacy but also on the immunization coverage, $C$, and the time needed to achieve such a coverage $\tau$. 


%

\begin{figure}[p]\centering
  \begin{subfigure}[h]{\textwidth} \centering
 \includegraphics[width=0.9\textwidth]{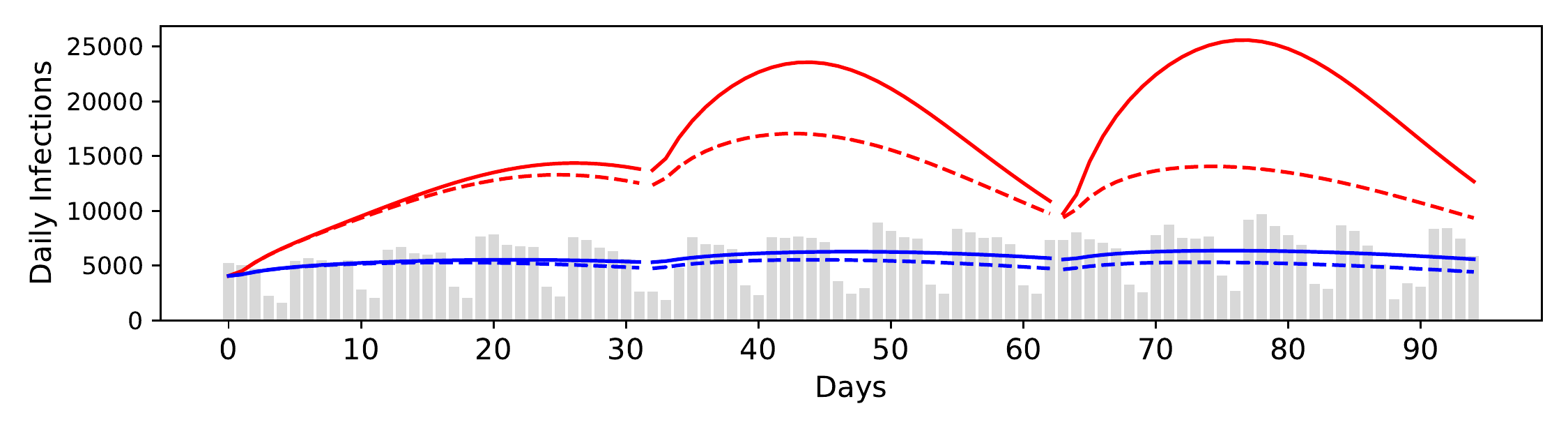}\caption{ }
 \includegraphics[width=0.9\textwidth]{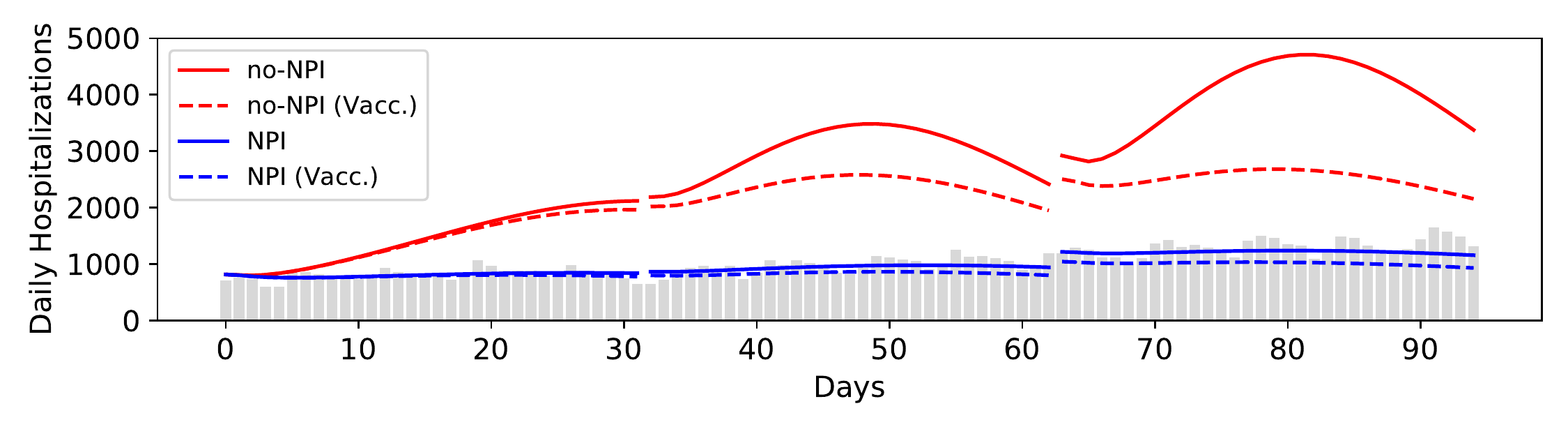}\caption{ }
  \includegraphics[width=0.9\textwidth]{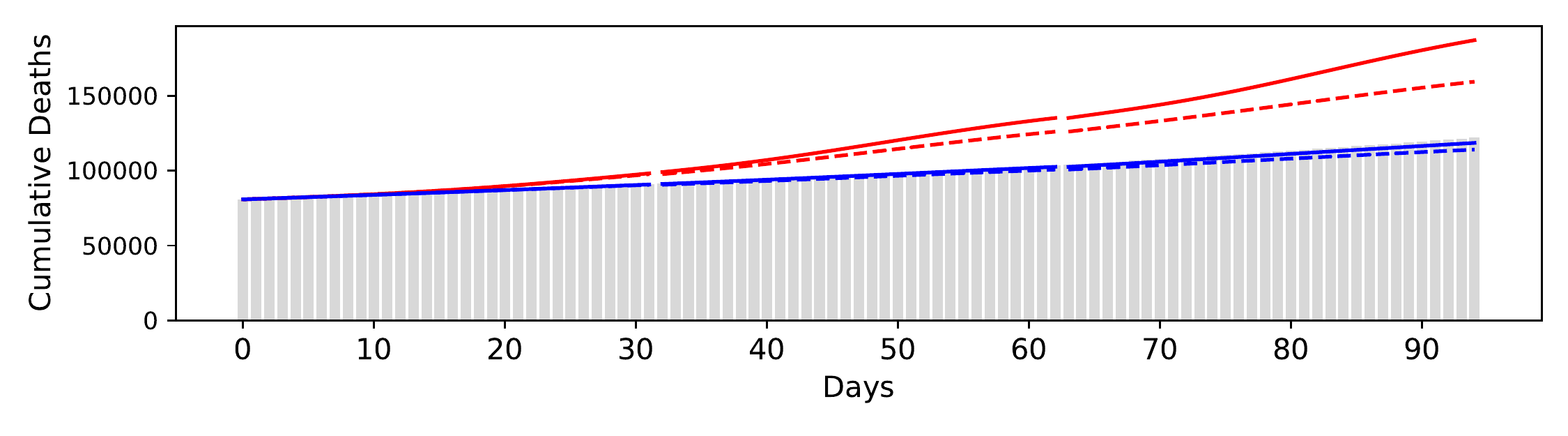}\caption{ }
 \end{subfigure}
\caption{Impact of a vaccination program on the number of (a) new daily infections, (b) hospitalizations, and (c) cumulative deaths (dashed lines). The no-NPIs scenario is shown in solid red lines and the scenario under NPIs implementation is shown in solid blue lines. The initial deployment of the vaccine starts on October 1. Vaccination parameters are as follows: coverage time $\tau=90$ days, vaccination coverage $C=0.20$, efficacy $\psi=0.90$, and symptomatic fraction $\tilde{p}=0.30$. Bar plots represent officially reported data.}
\label{fig4}
\end{figure}


The population-level impact of a vaccination program on the number of new daily infections, hospitalizations, and cumulative deaths is shown in Figure \ref{fig4} (a), (b), and (c), respectively. To have a better understanding of the effect of vaccination, we consider its interaction with the impact of implementing or partially lifting NPIs. The no-NPIs scenario, that is, the case in which no mitigation measures are implemented at all, is shown in solid red lines. On the other hand, the disease burden for the case in which NPIs are implemented is shown in solid blue lines (see Figure \ref{fig4}). Moreover, dashed lines correspond to the respective scenario considering the introduction of a vaccination program. Vaccination parameters are as follows: coverage time $\tau=90$ days, vaccination coverage $C=0.20$, efficacy $\psi=0.90$, and symptomatic fraction $\tilde{p}=0.30$. The vaccination rate $\phi$ is obtained from the approximation $1-\exp\left\lbrace -\phi\tau\right\rbrace =C$. The initial deployment of the vaccine starts on October 1, so we employ our estimated parameters for October, November, and December. The results shown in Figure \ref{fig4} point out that, in the absence of NPIs, the immunization coverage needed to control the pandemic is very high. 

\begin{figure}\centering

\begin{subfigure}[h]{0.3\textwidth} \centering
 \includegraphics[width=\textwidth]{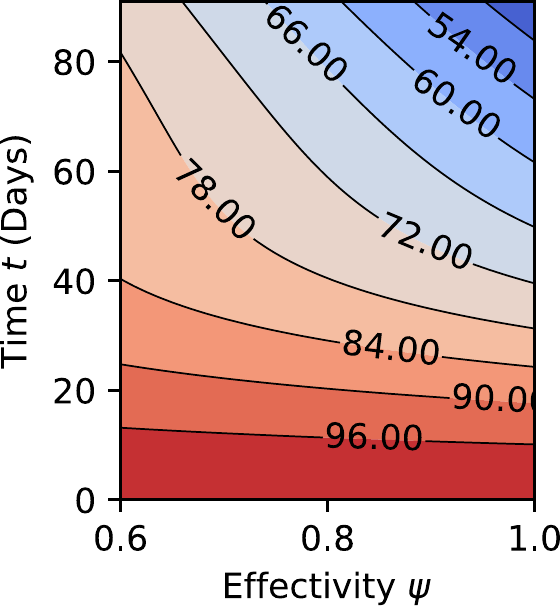}\caption{ }
 \end{subfigure}
 \begin{subfigure}[h]{0.3\textwidth} \centering
 \includegraphics[width=\textwidth]{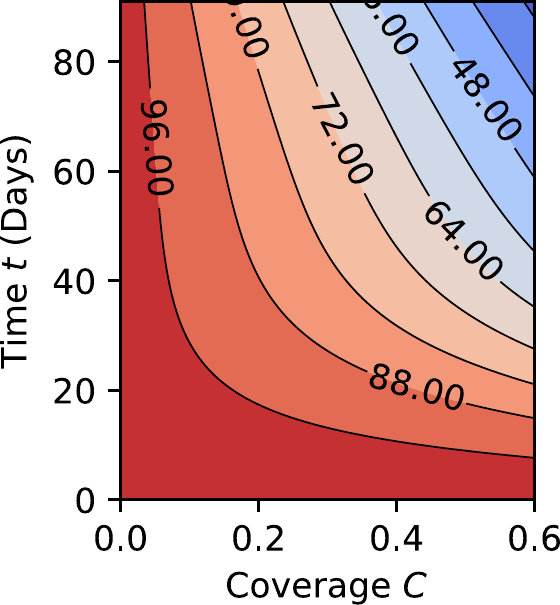}\caption{ }
  \end{subfigure}
  
  \begin{subfigure}[h]{0.3\textwidth} \centering
 \includegraphics[width=\textwidth]{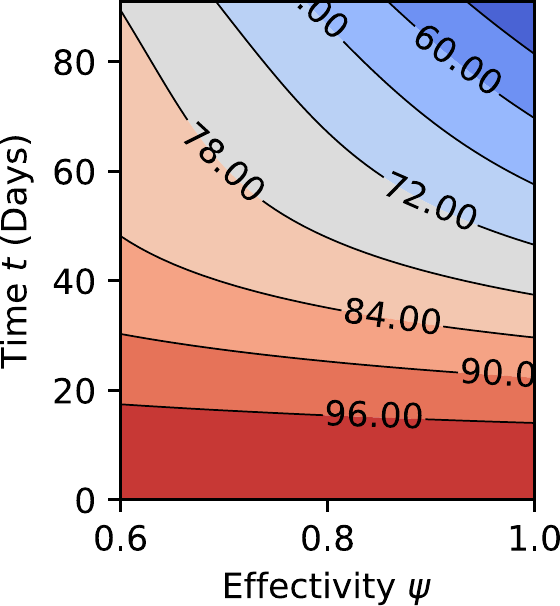}\caption{ }
 \end{subfigure}
 \begin{subfigure}[h]{0.3\textwidth} \centering
 \includegraphics[width=\textwidth]{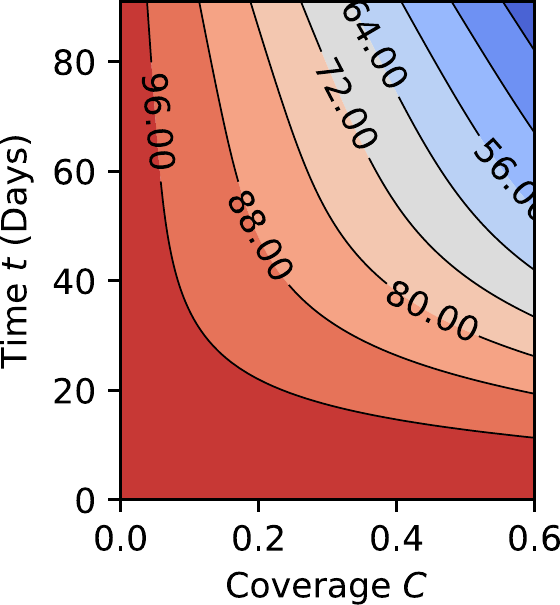}\caption{ }
  \end{subfigure}
  
  \begin{subfigure}[h]{0.3\textwidth} \centering
 \includegraphics[width=\textwidth]{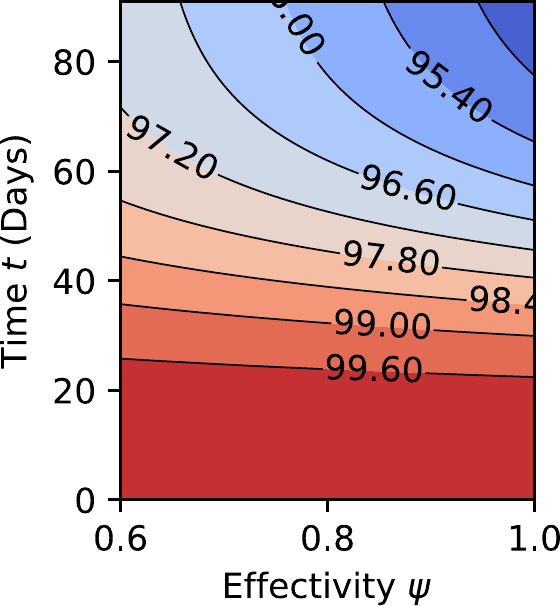}\caption{ }
 \end{subfigure}
 \begin{subfigure}[h]{0.3\textwidth} \centering
 \includegraphics[width=\textwidth]{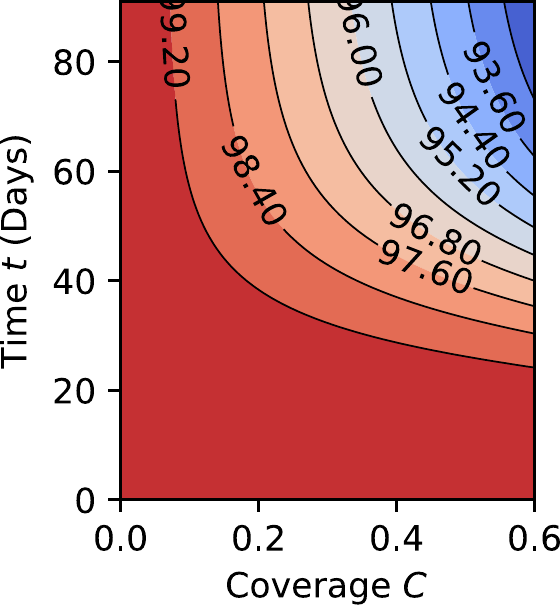}\caption{ }
  \end{subfigure}

\caption{Countour plots that show the percentage scale of the pandemic burden under vaccination with respect to the scenario without NPIs for (a)-(b) new daily infections, (c)-(d) new daily hospitalizations and (e)-(f) cumulative deaths. For all plots, vaccination starts on November 1. When fixed, the vaccine-associated parameters are coverage time $\tau = 90$ days, target coverage $C=40\%$, effectiveness $\psi = 0.90$ and symptomatic fraction $\tilde{p} = 0.40$.}
\label{fig5}
\end{figure}

We further explore the joint impact vaccine introduction and implementation of NPIs in Figure \ref{fig5}. The contour plots show the percentage scale of the pandemic burden under vaccination concerning the scenario without NPIs for (a)-(b) new daily infections, (c)-(d) new daily hospitalizations, and (e)-(f) cumulative deaths. In Figure \ref{fig5}, a value of $60\%$ means that the corresponding curve under vaccination is $60\%$ the value of the same curve under the scenario without NPIs. We explore the effects of varying vaccine efficacy $\psi$, and vaccination coverage $C$. We also explored the effect of varying the symptomatic fraction $\tilde{p}$, and we found that this parameter does not affect significantly the outcomes under study. On the other hand, the vaccination coverage $C$ has a greater effect in terms of reducing the disease burden. However, the immunization coverage needed to see a significant reduction in the number of infections should be close to 40\%.

\begin{figure}[tbp]\centering

\begin{subfigure}[h]{\textwidth} \centering
 \includegraphics[width=\textwidth]{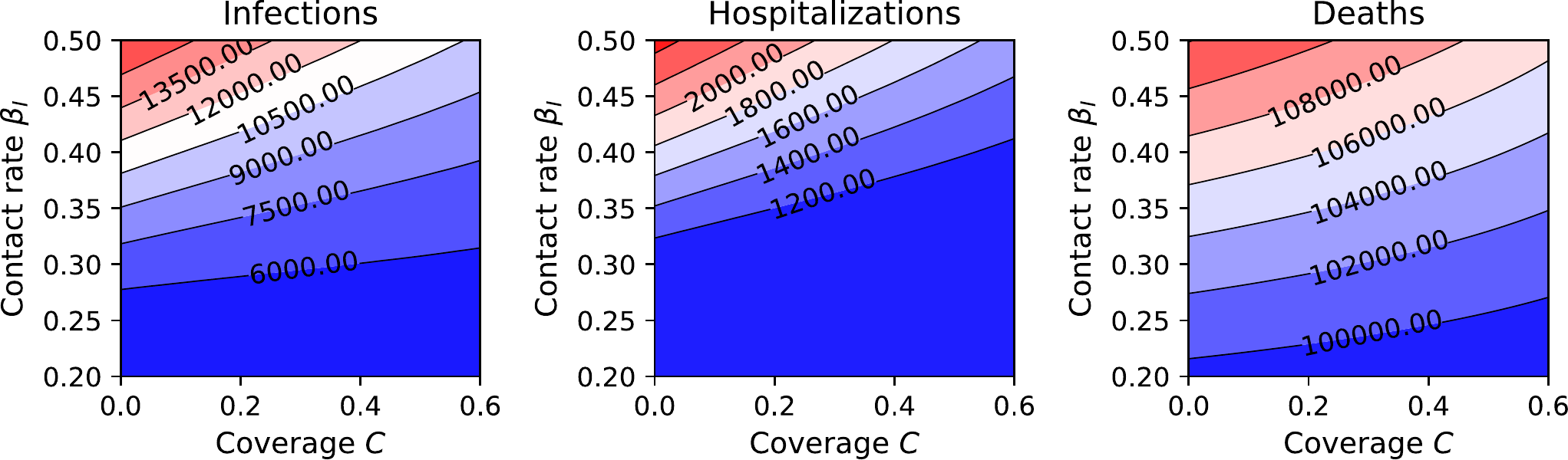}\caption{ }
  \includegraphics[width=\textwidth]{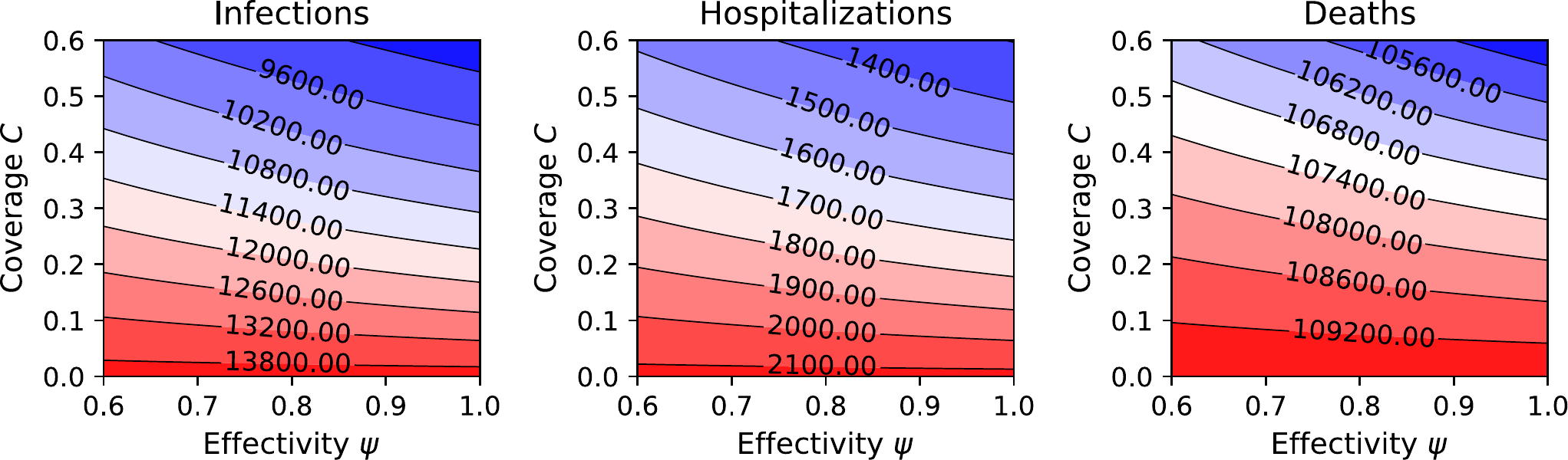}\caption{ }
 \end{subfigure}

\caption{Impact of vaccine-associated parameters on the maximum value of new daily infections, new daily hospitalizations and cumulative deaths for November. (a) The immunization coverage $C$ and the contact rate $\beta_{I}$ are varied (the rest of parameters are fixed $\psi = 0.90$, $\tilde{p}=0.40$ and $\tau = 30$ days). (b) The vaccine efficacy $\psi$ and coverage $C$ are varied (the rest of parameters are fixed $\tilde{p}=0.40$ and $\tau = 30$ days).}
\label{fig6}
\end{figure}

In Figure \ref{fig6}, we show the maximum value for new daily infections, new daily hospitalizations, and cumulative deaths in a window of 30 days predicted for a vaccination program starting on November 1 with variations in vaccine-associated parameters. In Figure \ref{fig6}(a), we explore the impact of vaccination target coverage, $C$, against the contact rate $\beta_I$. As expected, increasing coverage is a very effective way to control the pandemic. However, assuming realistically low coverage levels (close to ~10\% within the first weeks of vaccine introduction), we can observe that relaxing NPIs altogether with low vaccination coverage may dramatically increase the disease burden \cite{Abo2020Dec}. However, if the implementation of NPIs manages to maintain a low contact rate, the introduction of the vaccine improves notably the control of the disease burden.  In Figure \ref{fig6}(b), we show the effects of varying vaccine efficacy $\psi$ along with coverage $C$. The results imply that it may be more desirable to employ a vaccine with low efficacy but reach a high coverage than a vaccine with high effectiveness but low coverage levels \cite{Paltiel2020Nov}. 

%

\section{Discussion}
\label{sec:discussion}


Since the emergence of SARS-CoV-2, health authorities worldwide have implemented unprecedented mobility restrictions and other NPIs in an attempt to control the epidemic. Nevertheless, due to socio-economic repercussions, many countries decided to lift or at least partially relax such restrictions causing a worrisome rebound of the epidemic. Due to constant modifications of NPIs according to epidemiological risk factors, it is very difficult that the deterministic dynamics of a compartmental model may fit long-term data or multiple waves of the epidemic. Moreover, the forecast in the absence of data is limited to conditions remaining constant which is unrealistic due to the changes needed to reactivate the economy. To circumvent this situation, we consider a monthly parameter inference. This strategy allows us to observe and compare the progression of transmission parameters alongside the epidemic. Besides, Bayesian inference allowed us to calibrate key model features and evaluate the effects of NPIs given data. In particular, we used official epidemiological data at the early phase of the epidemic in Mexico, before lockdowns and NPIs implementation, to establish the effective contact rate without mitigation measures. This allowed us to assess the effects of NPIs in the following months of the epidemic. Our model simulations show that the number of daily infections, hospitalizations, and cumulative deaths would have presented at least a two-fold increase in the absence of NPIs. Hence, the implementation of NPIs in Mexico has been of paramount importance to reduce the transmission of SARS-CoV-2 and the morbidity caused by COVID-19 disease.

After model calibration, we explored how a vaccination program affects the control of the transmission dynamics in comparison with NPIs. The results suggest that vaccination alone is not enough to control disease spread if NPIs are abandoned prematurely and if the immunization coverage is low as expected in some countries or regions. 
In other words, lifting mitigation measures completely at the early stages of vaccination may lead to a dramatic increase in the disease burden. Therefore, even though mass vaccination programs have already started all around the world, mobility restrictions and other NPIs are still of principal importance in the control of the COVID-19 pandemic. Furthermore, the simulations also suggest that it may be more desirable to employ a vaccine with low efficacy but reach a high coverage than a vaccine with high effectiveness but low coverage levels. This supports the hypothesis that delaying the second dose and prioritizing giving the first doses of vaccine to more individuals will be more optimal to mitigate the COVID-19 epidemic.

\section*{Acknowledgements}

Fernando Salda\~na acknowledges support from DGAPA-PAPIIT-UNAM grant IV 100220 (proyecto especial COVID-19). Ariel Camacho thanks SEP-SES-PRODEP-UABC for his Postdoctoral Fellowship support.  We thank Dr. Jorge X. Velasco-Hern\'andez for helpful discussions on a previous version of this work.

\pagebreak

\bibliographystyle{apalike}
\bibliography{references}

\begin{thebibliography}{}

\bibitem[Abo and Smith?, 2020]{Abo2020Dec}
Abo, S. M.~C. and Smith?, S.~R. (2020).
\newblock {Is a COVID-19 Vaccine Likely to Make Things Worse?}
\newblock {\em Vaccines}, 8(4):761.

\bibitem[Acu{\~n}a-Zegarra et~al., 2020]{acuna2020modeling}
Acu{\~n}a-Zegarra, M.~A., Santana-Cibrian, M., and Velasco-Hernandez, J.~X.
  (2020).
\newblock Modeling behavioral change and covid-19 containment in mexico: A
  trade-off between lockdown and compliance.
\newblock {\em Mathematical Biosciences}, page 108370.

\bibitem[Aguiar et~al., 2020]{aguiar2020modelling}
Aguiar, M., Ortuondo, E.~M., Van-Dierdonck, J.~B., Mar, J., and Stollenwerk, N.
  (2020).
\newblock Modelling covid 19 in the basque country from introduction to control
  measure response.
\newblock {\em Scientific reports}, 10(1):1--16.

\bibitem[Anirudh, 2020]{anirudh2020mathematical}
Anirudh, A. (2020).
\newblock Mathematical modeling and the transmission dynamics in predicting the
  covid-19-what next in combating the pandemic.
\newblock {\em Infectious Disease Modelling}, 5:366--374.

\bibitem[Bertozzi et~al., 2020]{bertozzi2020challenges}
Bertozzi, A.~L., Franco, E., Mohler, G., Short, M.~B., and Sledge, D. (2020).
\newblock The challenges of modeling and forecasting the spread of covid-19.
\newblock {\em Proceedings of the National Academy of Sciences},
  117(29):16732--16738.

\bibitem[Byrne et~al., 2020]{byrne2020inferred}
Byrne, A.~W., McEvoy, D., Collins, A., Hunt, K., Casey, M., Barber, A., Butler,
  F., Griffin, J., Lane, E., McAloon, C., et~al. (2020).
\newblock Inferred duration of infectious period of sars-cov-2: rapid scoping
  review and analysis of available evidence for asymptomatic and symptomatic
  covid-19 cases.
\newblock {\em medRxiv}.

\bibitem[Capistran et~al., 2021]{capistran2020forecasting}
Capistran, M.~A., Capella, A., and Christen, J.~A. (2021).
\newblock Forecasting hospital demand in metropolitan areas during the current
  covid-19 pandemic and estimates of lockdown-induced 2nd waves.
\newblock {\em PloS one}, 16(1):e0245669.

\bibitem[Christen et~al., 2010]{christen2010general}
Christen, J.~A., Fox, C., et~al. (2010).
\newblock A general purpose sampling algorithm for continuous distributions
  (the t-walk).
\newblock {\em Bayesian Analysis}, 5(2):263--281.

\bibitem[Delamater et~al., 2019]{delamater2019complexity}
Delamater, P.~L., Street, E.~J., Leslie, T.~F., Yang, Y.~T., and Jacobsen,
  K.~H. (2019).
\newblock Complexity of the basic reproduction number (r0).
\newblock {\em Emerging infectious diseases}, 25(1):1.

\bibitem[Diekmann et~al., 1990]{diekmann1990definition}
Diekmann, O., Heesterbeek, J. A.~P., and Metz, J.~A. (1990).
\newblock On the definition and the computation of the basic reproduction ratio
  r 0 in models for infectious diseases in heterogeneous populations.
\newblock {\em Journal of mathematical biology}, 28(4):365--382.

\bibitem[Hernandez-Vargas and Velasco-Hernandez, 2020]{hernandez2020host}
Hernandez-Vargas, E.~A. and Velasco-Hernandez, J.~X. (2020).
\newblock In-host mathematical modelling of covid-19 in humans.
\newblock {\em Annual reviews in control}.

\bibitem[Hethcote, 2000]{hethcote2000mathematics}
Hethcote, H.~W. (2000).
\newblock The mathematics of infectious diseases.
\newblock {\em SIAM review}, 42(4):599--653.

\bibitem[Ku et~al., 2020]{ku2020epidemiological}
Ku, C.~C., Ng, T.-C., and Lin, H.-H. (2020).
\newblock Epidemiological benchmarks of the covid-19 outbreak control in china
  after wuhan’s lockdown: a modelling study with an empirical approach.
\newblock {\em Available at SSRN 3544127}.

\bibitem[Maier and Brockmann, 2020]{maier2020effective}
Maier, B.~F. and Brockmann, D. (2020).
\newblock Effective containment explains subexponential growth in recent
  confirmed covid-19 cases in china.
\newblock {\em Science}, 368(6492):742--746.

\bibitem[Mandel and Veetil, 2020]{mandel2020economic}
Mandel, A. and Veetil, V. (2020).
\newblock The economic cost of covid lockdowns: An out-of-equilibrium analysis.
\newblock {\em Economics of Disasters and Climate Change}, 4(3):431--451.

\bibitem[Mena et~al., 2020]{mena2020using}
Mena, R.~H., Velasco-Hernandez, J.~X., Mantilla-Beniers, N.~B.,
  Carranco-Sapi{\'e}ns, G.~A., Benet, L., Boyer, D., and Castillo, I.~P.
  (2020).
\newblock Using the posterior predictive distribution to analyse epidemic
  models: Covid-19 in mexico city.
\newblock {\em arXiv preprint arXiv:2005.02294}.

\bibitem[Nogrady, 2020]{nogrady2020data}
Nogrady, B. (2020).
\newblock What the data say about asymptomatic covid infections.
\newblock {\em Nature}.

\bibitem[Oran and Topol, ]{oranproportion}
Oran, D.~P. and Topol, E.~J.
\newblock The proportion of sars-cov-2 infections that are asymptomatic: A
  systematic review.
\newblock {\em Annals of Internal Medicine}.

\bibitem[Paltiel et~al., 2020]{Paltiel2020Nov}
Paltiel, A.~D., Schwartz, J.~L., Zheng, A., and Walensky, R.~P. (2020).
\newblock {Clinical Outcomes Of A COVID-19 Vaccine: Implementation Over
  Efficacy}.
\newblock {\em Health Aff.}

\bibitem[Petropoulos and Makridakis, 2020]{petropoulos2020forecasting}
Petropoulos, F. and Makridakis, S. (2020).
\newblock Forecasting the novel coronavirus covid-19.
\newblock {\em PloS one}, 15(3):e0231236.

\bibitem[Ponciano and Capistr{\'a}n, 2011]{ponciano2011first}
Ponciano, J.~M. and Capistr{\'a}n, M.~A. (2011).
\newblock First principles modeling of nonlinear incidence rates in seasonal
  epidemics.
\newblock {\em PLoS Comput Biol}, 7(2):e1001079.

\bibitem[Ribeiro et~al., 2020]{ribeiro2020short}
Ribeiro, M. H. D.~M., da~Silva, R.~G., Mariani, V.~C., and dos Santos~Coelho,
  L. (2020).
\newblock Short-term forecasting covid-19 cumulative confirmed cases:
  Perspectives for brazil.
\newblock {\em Chaos, Solitons \& Fractals}, 135:109853.

\bibitem[Roda et~al., 2020]{roda2020difficult}
Roda, W.~C., Varughese, M.~B., Han, D., and Li, M.~Y. (2020).
\newblock Why is it difficult to accurately predict the covid-19 epidemic?
\newblock {\em Infectious Disease Modelling}, 5:271--281.

\bibitem[Salda{\~n}a et~al., 2020]{saldana2020modeling}
Salda{\~n}a, F., Flores-Arguedas, H., Camacho-Guti{\'e}rrez, J.~A., and
  Barradas, I. (2020).
\newblock Modeling the transmission dynamics and the impact of the control
  interventions for the covid-19 epidemic outbreak.

\bibitem[Saldana and Velasco-Hernandez, 2020]{saldana2020trade}
Saldana, F. and Velasco-Hernandez, J.~X. (2020).
\newblock The trade-off between mobility and vaccination for covid-19 control:
  a metapopulation modeling approach.
\newblock {\em medRxiv}.

\bibitem[Salud, 2020]{datosSS}
Salud, S. (2020).
\newblock Datos covid-19 m\'exico.
\newblock https://datos.covid-19.conacyt.mx/.
\newblock Accessed 12-11-2020.

\bibitem[Santamar{\'\i}a-Holek and Casta{\~n}o, 2020]{santamaria2020possible}
Santamar{\'\i}a-Holek, I. and Casta{\~n}o, V. (2020).
\newblock Possible fates of the spread of sars-cov-2 in the mexican context.
\newblock {\em Royal Society open science}, 7(9):200886.

\bibitem[Santana-Cibrian et~al., 2020]{santana2020lifting}
Santana-Cibrian, M., Acu{\~n}a-Zegarra, M.~A., and Velasco-Hernandez, J.~X.
  (2020).
\newblock Lifting mobility restrictions and the effect of superspreading events
  on the short-term dynamics of covid-19.
\newblock {\em Mathematical Biosciences and Engineering}, 17(5):6240--6258.

\bibitem[Sarkar et~al., 2020]{sarkar2020modeling}
Sarkar, K., Khajanchi, S., and Nieto, J.~J. (2020).
\newblock Modeling and forecasting the covid-19 pandemic in india.
\newblock {\em Chaos, Solitons \& Fractals}, 139:110049.

\bibitem[Su et~al., 2021]{su2021vaccines}
Su, Z., Wen, J., McDonnell, D., Goh, E., Li, X., {\v{S}}egalo, S., Ahmad, J.,
  Cheshmehzangi, A., and Xiang, Y.-T. (2021).
\newblock Vaccines are not yet a silver bullet: The imperative of continued
  communication about the importance of covid-19 safety measures.
\newblock {\em Brain, Behavior, \& Immunity-Health}, page 100204.

\bibitem[Torrealba-Rodriguez et~al., 2020]{torrealba2020modeling}
Torrealba-Rodriguez, O., Conde-Guti{\'e}rrez, R., and Hern{\'a}ndez-Javier, A.
  (2020).
\newblock Modeling and prediction of covid-19 in mexico applying mathematical
  and computational models.
\newblock {\em Chaos, Solitons \& Fractals}, 138:109946.

\bibitem[Van~den Driessche and Watmough, 2002]{van2002reproduction}
Van~den Driessche, P. and Watmough, J. (2002).
\newblock Reproduction numbers and sub-threshold endemic equilibria for
  compartmental models of disease transmission.
\newblock {\em Mathematical biosciences}, 180(1-2):29--48.

\bibitem[Zhao and Feng, 2020]{zhao2020staggered}
Zhao, H. and Feng, Z. (2020).
\newblock Staggered release policies for covid-19 control: Costs and benefits
  of relaxing restrictions by age and risk.
\newblock {\em Mathematical biosciences}, 326:108405.

\end{thebibliography}

\pagebreak
\begin{center}
\textbf{\large Supplementary Material: Estimating the impact of non-pharmaceutical interventions and vaccination on the progress of the COVID-19 epidemic in Mexico: a mathematical approach }
\end{center}
\setcounter{equation}{0}
\setcounter{figure}{0}
\setcounter{table}{0}
\setcounter{page}{1}
\setcounter{section}{0}
\makeatletter
\renewcommand{\thesection}{S-\Roman{section}}
\renewcommand{\theequation}{S\arabic{equation}}
\renewcommand{\thefigure}{S\arabic{figure}}
\renewcommand{\bibnumfmt}[1]{[S#1]}
\renewcommand{\citenumfont}[1]{S#1}

\section{Mathematical properties of the model}\label{sec:appendixA}

The biologically feasible region for model \eqref{model1} is
\begin{equation*}
\Omega=\left\lbrace (S, V, E, \tilde{E}, A, I, R, D)\in \mathbb{R}^{8}_{+}: S+V+E+\tilde{E}+A+I+R+D=N\right\rbrace.
\end{equation*}
Clearly, the region $\Omega$ is positively-invariant, that is, for a well-defined initial condition that starts in $\Omega$, the solution remains in $\Omega$ for all $t>0$. Therefore, the model is  both epidemiologically and mathematically well posed \citep{hethcote2000mathematics}.

System \eqref{model1} presents a continuum of disease-free equilibria of the form 
\begin{equation*}
E_{0}=(S, V, E, \tilde{E}, A, I, H, R, D)=(S^{*}, V^{*}, 0, 0, 0, 0, 0, 0, 0),
\end{equation*}
where $S^{*}$, and $V^{*}$ are the proportions of non-vaccinated and vaccinated susceptible at the initial time. An straightforward computation allow us to obtain the basic reproduction number $\mathcal{R}_{0}= (1-p)\beta_{A}/\gamma_{A} + p \beta_{I}/(\gamma+\eta +\mu)$. Note that, by definition, $\mathcal{R}_{0}$ assumes a fully susceptible population and, hence, it cannot be modified through vaccination campaigns. To examine the effects of vaccination and non-pharmaceutical interventions, the more appropriate measure to use is the effective reproduction number $\mathcal{R}_{e}$ \citep{delamater2019complexity}. 

We obtain $\mathcal{R}_{e}$ taking a next-generation approach \citep{diekmann1990definition}. The matrix of new infections $\mathbf{F}$ and the matrix of changes in the infection status $\mathbf{V}$ are given by:
\begin{equation*}
\mathbf{F}= 
\begin{bmatrix}
0 & 0 & \beta_{A}S^{*}/N^{*} & \beta_{I}S^{*}/N^{*} \\
0 & 0 & \beta_{A}(1-\psi)V^{*}/N^{*} & \beta_{I}(1-\psi)V^{*}/N^{*} \\
0 & 0 & 0 & 0 \\
0 & 0 & 0 & 0 \\
\end{bmatrix}
\end{equation*}
and
\begin{equation*}
\mathbf{V}= 
\begin{bmatrix}
k       & 0               & 0 & 0 \\
0       & k               & 0 & 0 \\
-(1-p)k & -(1-\tilde{p})k & \gamma_{A} & 0 \\
-pk     & -\tilde{p}k     & 0 & \gamma + \eta + \nu \\
\end{bmatrix}.
\end{equation*}
The next-generation matrix is $\mathbf{K}=\mathbf{F}\mathbf{V}^{-1}$ and the effective reproduction number is the spectral radius $\mathcal{R}_{e}=\rho(\mathbf{K})$. If the symptomatic fractions satisfy $p=\tilde{p}$, it is easy to see that $\mathcal{R}_{e}=(S^{*}/N^{*}+(1-\psi)V^{*}/N^{*})\mathcal{R}_{0}$, and in the absence of vaccination $\mathcal{R}_{e}=(S^{*}/N^{*})\mathcal{R}_{0}$. Therefore, we approximate the time-varying effective reproduction number as the product of the proportion of the susceptible among the effective population size at the beginning of each month, i.e. $\mathcal{R}_{e}=(S_0/N_e)\mathcal{R}_{0}$ (see also \citep{ku2020epidemiological}). Let us note that using an explicit formula for $\mathcal{R}_{e}$ based on the dynamic of the model means that the same assumptions are taken into account. That is, the value of $\mathcal{R}_{e}$ includes the postulates about the non observed dynamic. Although it may produce discrepancies with other ways of calculating it, the monthly computation of $\mathcal{R}_{e}$ allows us to assess the efficacy of NPIs implemented. In section \ref{sec:parameters}, we obtain a predictive marginal for $\mathcal{R}_{e}$ based on the marginal posterior for $S_0$ and the predictive marginal of $R_0$.  As a consequence of the Theorem 2 in \cite{van2002reproduction}, we establish the following result regarding the local stability of the disease-free equilibrium.

\begin{theorem}
The continuum of disease-free equilibrium of system \eqref{model1} given by $E_{0}$ is locally asymptotically stable if the effective reproduction number satisfies $\mathcal{R}_{e}<1$ and unstable if $\mathcal{R}_{e}>1$.
\end{theorem}

\section{Bayesian Inference}\label{sec:appendixB}
To perform the parameter inference, we first retrieved some of the model parameters from the literature and epidemiological data on COVID-19. The incubation period of COVID-19 is on average 5-6 days, but can be as long as 14 days, we choose the average estimation $1/k=5.1$ days \cite{acuna2020modeling}. There is substantial variation in the estimates for the infectious period. Some studies provided an approximate median infectious period for asymptomatic cases of $6.5 - 9.5$ days and a potential maximum infectious period of 18 days in the symptomatic cases \cite{byrne2020inferred}; here, we assume $1/\gamma_{A}=1/\gamma=7.0$ days. Mortality data shows that most deaths come from hospitalized patients \cite{datosSS}. Hence, we assume that $m = 10 \mu$, so the mortality rate is higher in the hospitalized class than in the symptomatic class. Recent studies suggest that at least one third of SARS-CoV-2 infections are asymptomatic \cite{oranproportion}, thus, we postulate the value $p = 0.5$. With respect to the asymptomatic effective contact rate, we assume the relation $\beta_A = 2 \beta_I$ since symptomatic individuals are the ones that develop severe conditions and therefore, on average, are expected to have reduced mobility compared with asymptomatic cases. We remark that some studies have found that completely asymptomatic individuals will transmit the virus to significantly fewer people than a symptomatic case \cite{nogrady2020data}. However, in our model the asymptomatic class $A$ includes both, completely asymptomatic and mild infections who do not attend a medical center.

The initial value of the state variables (corresponding to March, 2020) are fixed as follows: $A(0)=20$, and  $R(0) = H(0) = D(0) = 0$. The rest of initial conditions, $S(0), E(0),$ and $I(0)$, are included in the inference process. We assume the following model for new infections and new hospitalizations data $y_i$ 
\begin{equation}
y_i \sim \text{Poisson} \left(\mathcal{I}_j(\mathbf{x})\right), \qquad  i= 0, \ldots, k
\end{equation}
where $\mathcal{I}_j(\mathbf{x})$ denote the predicted number of new cases between times $j-1$ and $j$, see \cite{ponciano2011first}.
For the case of the infections in our model, 
\begin{equation}
\label{eq:incidence_infec}
\mathcal{I}^I_j(\mathbf{x})=\int_{t_{j-1}}^{t_j}pk E(\mathbf{x}) dt
\end{equation}
with $E$ the exposed individuals given by system \eqref{model1} at time $t_i$ and $\mathbf{x}$ the vector of parameters to estimate. By assuming independence on the observations, the likelihood function $\mathcal{L}(\mathbf{x})$ satisfies:
\begin{equation}
\label{eq:likelihood}
\mathcal{L}(\mathbf{x})\propto \prod_{j=0}^{j=k} \dfrac{e^{-\mathcal{I}^I_j(\mathbf{x})}(\mathcal{I}^I_j(\mathbf{x}))^{y_j}}{y_j!}.
\end{equation}
Analogously, we consider
\begin{equation}
\label{eq:incidence_hosp}
\mathcal{I}^H_j(\mathbf{x})=\int_{t_{j-1}}^{t_j}\eta I(\mathbf{x}) dt
\end{equation}
the incidence for the new hospitalizations. In the case of deaths, as at the beginning of the epidemic there were only a few, we consider cumulative cases and propose a Gaussian likelihood to fit this data.

\begin{figure}
\centering
\includegraphics[width=\textwidth]{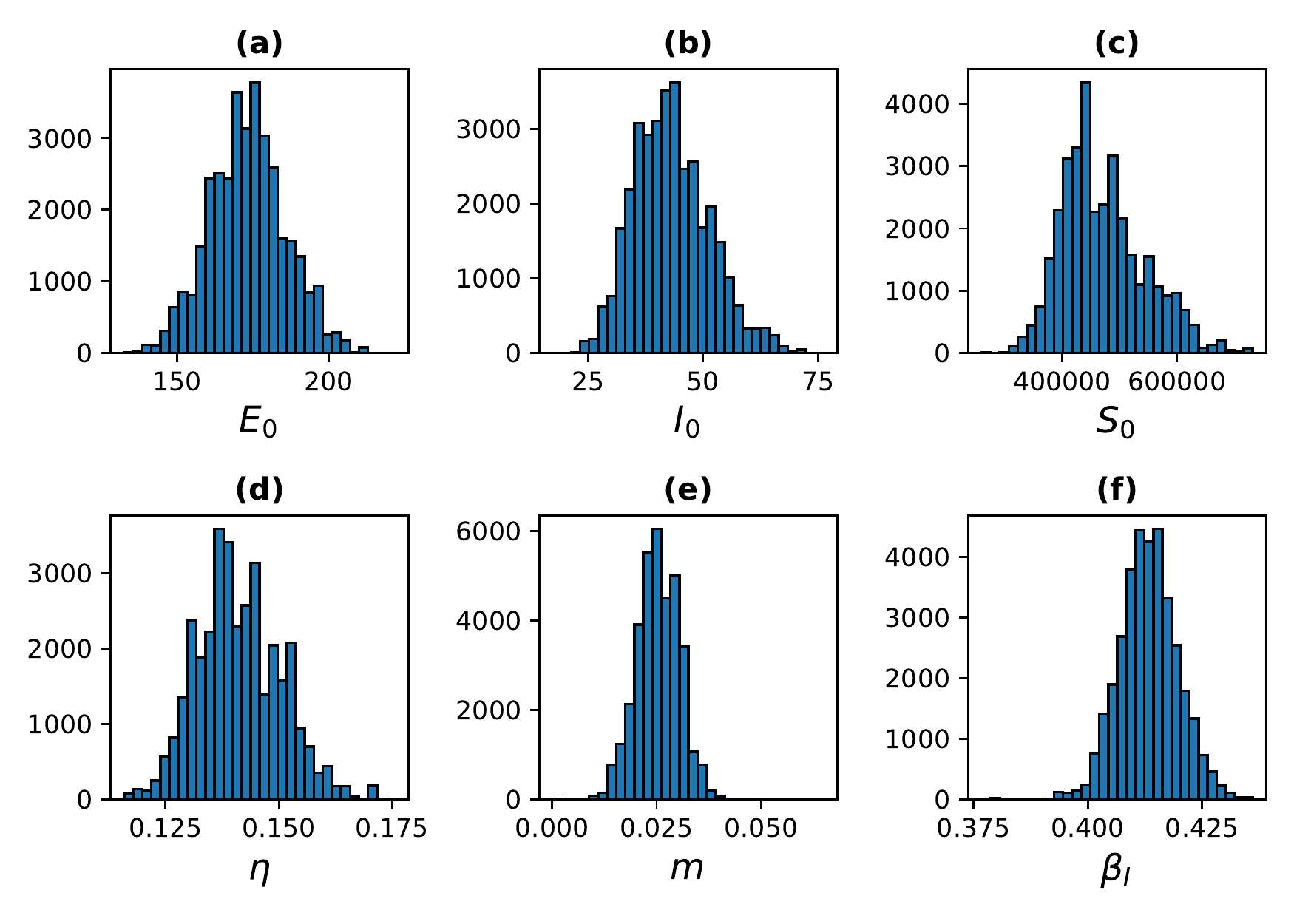}
\caption{Marginal posterior for the parameters in $\mathbf{x}$: \textbf{(a)} $E_0$, \textbf{(b)} $I_0$, \textbf{(c)} $S_0$, \textbf{(d)} $\eta$, \textbf{(e)} $m$ and \textbf{(f)} $\beta_I$.}
\label{fig: marginal_posterior1}
\end{figure}

We define by $\pi_0(\mathbf{x})$ the prior distribution for $\mathbf{x}$. We assume independence of the parameters, hence
\begin{equation}
\pi_0(\mathbf{x})=\pi^0_0(S_0)\pi^1_0(\beta_I)\pi^2_0(E_0)\pi^3_0(I_0)\pi^4_0(\eta)\pi^5_0(m).
\end{equation}
We consider gamma distributions for each one. Recall that the gamma distribution is denoted by $\Gamma (\alpha, \beta)$ with $\alpha$ the shape parameter and $\beta$ the inverse scale parameter. If $Z \sim \Gamma (\alpha, \beta)$ then $\mathbb{E}[Z]=\alpha / \beta$ and $\mathbb{V}ar[Z]=\alpha /{\beta^2}$. We propose 
\begin{equation}
\begin{array}{ccc}
S_0  \sim \Gamma (a, b), &
\beta_{I} \sim \Gamma (1, 1), &
E_0  \sim \Gamma (20, 1), \\
I_0  \sim \Gamma (20, 1), &
\eta  \sim \Gamma (0.1, 1), &
m  \sim \Gamma (0.1 ,1 )
\end{array}
\label{eq:prior_distributions}
\end{equation}
where $a$ and $b$ are chosen such that $\mathbb{E}[S_0]=5.0e5$ with an standard deviation of 7.0e4. As the national lockdown was implemented in Mexico around March 25, we use data from March to estimate the baseline transmission rate before the implementation of the lockdown and other NPIs. The posterior marginals of the parameters in $\mathbf{x}$ are obtained using the t-walk \cite{christen2010general} and are shown in Figure \ref{fig: marginal_posterior1}. It is natural to expect a lower contact rate for April and May as a consequence of the lockdown. Since there was no vaccination in the early phase of the pandemic, compartments $V$ and $\tilde{E}$ are not considered in the inference. 

The inference allows us to obtain a predictive marginal for state variables $(S,E,A,I,H,R,D)$ at the final date of March. These distributions will provide information to perform the inference in the next month. The mean of these values will be used as initial condition of the model for the quantities $A,H,R,D$ and the marginal predictive of $S,E,$ and $I$ at March 31 will be used as prior distributions for the inference of April. This process is repeated to perform a parameter inference in a monthly way from April to December (see Figure \ref{fig1}).



\end{document}